\documentclass[a4paper]{article}
\usepackage[ansinew]{inputenc}
\usepackage{times}
\usepackage[T1]{fontenc}
\usepackage{graphicx}
\usepackage{geometry}
\usepackage{amssymb}
\usepackage{amsmath}

\usepackage{rotating}
\newcommand{\R}{\mathbb{R} }
\usepackage{tcolorbox}
\colorlet{shadecolor}{gray!25}
\usepackage{subfig}
\usepackage{float}
\newcommand{\RED}[1]{\color{red}{#1}\color{black}\Large\bfseries}

\newcommand{\Green}[1]{\textcolor{green}{#1}}
\usepackage[edges]{forest}
\usepackage[]{mdframed}
\usetikzlibrary{arrows.meta}
\usepackage[normalem]{ulem} %emphasize weiterhin kursiv
\usepackage {ulem} %\emph{Text}: Text w
\usepackage{colortbl}

\author{\normalsize Ludwig A. Hothorn\\ 
\footnotesize Im Grund 12, D-31867 Lauenau, Germany (e-mail:ludwig@hothorn.de)\\ 
Leibniz University Hannover (retired)}
\title{Bartholomew's $\bar{E}_k^2$ trend test - approximated by a multiple contrast test} 

\begin{document}
\maketitle
\begin{abstract}
Bartholomew's trend test belongs to the broad class of isotonic regression models, specifically with a single qualitative factor, e.g. dose levels.
Using the approximation of the ANOVA F-test by the maximum contrast test against grand mean and pool-adjacent-violator estimates under order restriction, an easier to use approximation is proposed. 
\end{abstract}

%%%%%%%%%%%%%%%%%%%%%%%%%%%%%%%%%%%%%%%%%%%%%%%%%%%%%%%%%%%%%%%%%%
\section{Bartholomew's trend test}

Bartholomew's $\bar{E}_k^2$ test \cite{bartholomew1959} for the simple ordered alternative $H_1: \mu_1\leq \mu_2\leq ...\leq \mu_k | \mu_1<\mu_k$ in a fully randomised one-way layout with $N(\mu_i, \sigma^2)$ ($i=1,... ,k$) is a modification of the usual ANOVA F-test (which is sensitive to any heterogeneity between qualitative treatment levels), replacing the arithmetic mean estimators $\bar{x_i}$ by the pool-adjacent-violators under order restriction (PAVA) estimates $\hat{\mu_i}$. This test belongs to the class of isotonic regression models, but for a single quantitative factor and not for a regression covariate or even dependent time series data (see the discussion for dependent data \cite{lei1995}). However, its use is severely limited by the lack of software (a permutative version is available in the CRAN package \textit{IsoGene} \cite{Lin2015}). It is limited to a simple one-way design with homogeneous variances and balanced designs.\
In the following, an approximation is derived based on the analogous approximation of the ANOVA-F test by multiple contrast tests (MCT) compared to the overall mean ($MCT_{GrandMean}$) according to \cite{Konietschke2013} and the use of the PAVA estimators instead of the arithmetic means. For k-sample comparisons, global tests can be formulated by the sum of the quadratic deviations from the overall mean, e.g. ANOVA F-test or $\bar{E}_k^2$ test, or by maxT tests for linear forms vs. overall mean, i.e. multiple contrasts (or their min-p standardisations). The $MCT_{GrandMean}^E$ trend test proposed here is based on maximum likelihood estimators under order restriction $\hat{\mu_i}$ (using PAVA) and the multiple contrast test against the overall mean. 

%%%%%%%%%%%%%%%%%%%%%%%%%%%%%%%%%%%%%%%%%%%%%%%%%%%%%%%%%%%%%%%%%%%%%%%%%%%%%
\subsection{An approximation: the $MCT_{homog}^E$ trend test}

An approximation of unspecific multiple contrast tests was already described \cite{miwa1998, miwa2000}. Here, a MCT against overall mean provides either simultaneous confidence intervals or compatible p-values as well as a global p-value by means of the min-p approach \cite{Pallmann2016}. MCT is an union-intersection test $MCT_{GrandMean}=max(t_1,...,t_\xi)$ based on the single contrasts\\
  $	t_{Contrast}=\sum_{i=0}^k c_i\hat{x}_i/S \sqrt{\sum_i^k c_i^2/n_i}$ where $c_i^q$ are the contrast coefficients. The contrast coefficients $c_i^q$ can be obtained easily (here didactically simplified for $i=3$ treatment groups $T_i$ in a balanced design):

        \begin{table}[H]
				\centering
        \begin{tabular}{ c  c c r c c c c }
         $c_i$ & $T_1$ & $T_2$ & $T_3$ \\ \hline
        $c_a$ & -1 & 1/2   & 1/2   \\
        $c_b$ & 1/2 & -1 & 1/2  \\
				$c_b$ & 1/2 & 1/2 & -1  \\
        \end{tabular}
				\caption{Grand Mean comparisons contrasts for k=3 treatment levels} 
				\label{tab:con}
				\end{table}
For general unbalanced one-way layouts the related contrast coefficients $c_i^q$ are available in the CRAN package \textit{multcomp} by means of the function \textit{contrMat(n, type = "GrandMean")} \cite{Hothorn2008}. \\
The above underlying maxT-test is a scalar test. Its multiple test statistics $(t_1,...,t_\xi)'$ follows jointly a $\xi$-variate $t$-distribution with the common degree of freedom $df$ and correlation matrix  $\R$ ($\R=f(c_{ij},n_i)$) based on a common variance estimator $S$ and common $df$. Both two-sided hypotheses as well as one-sided hypotheses can be formulated. Minor size and/or power problems arise for designs with rather small $n_i$ because most of the proposed generalizations are asymptotic approaches only \cite{Hothorn2008}. \\
Although there are a few examples where either an increasing or decreasing trend is tested, i.e. by means of a two-sided trend test. The majority of test for trend generally implies a one-sided hypothesis formulation. The hierarchy of hypothesis formulation in the k-sample case is: i) without restriction (like the ANOVA F-test) $\prec$ ii) one-sided comparisons (like the Dunnett test) $\prec$ iii) increasing trend (like the Armitage trend test for tumor incidences \cite{Armitage1955}). From this point of view, the $\bar{E}^2$ test is inconsistent: the quadratic test statistic is formulated two-sided, but the pool-adjacent-violators algorithm (PAVA) requires the directional specification of either increasing or decreasing beforehand. PAVA assumes an homogeneous global variance estimator $\sigma$ an a-priori defined direction (increase or decrease) and the weights $w_i=n_i$, available in the CRAN package \textit{Iso} \cite{Turner2023}. It is assumed that we have prior information which specifies restrictions of the form $\mu_i>\mu_j $  between some or all of $\mu's$ unknown with weights $n_i/\sigma$ (assuming homogeneous $\sigma$).

%%%%%%%%%%%%%%%%%%%%%%%%%%%%%%%%%%%%%%%%%%%%%%%%%%%%%%%%%%%%%%%%%%%%%
\subsubsection{Extensions}
Several extensions of the $MCT_{homog}^E$ trend test are straightforward. First, both 2-sided or 1-sided test versions are available. However, the 1-sided test is dominating, since the restriction of a general heterogeneity alternative to trend without an a-priori definition of the direction gives rarely sense. Second, the $MCT_{homog}^E$ trend test is available for any unbalanced design. Third, since variance homogeneity is a strict assumption and in several real data heterogeneous variances occur, a related modified test $MCT_{heterog}^E$ is available (see chapter \ref{heto}). Fourth, multiple contrast tests belong to the ANOVA-type models in the generalize linear mixed models (i.e. a qualitative factor is modeled). Therefore related asymptotic generalizations for e.g. proportions (see related approaches in \cite{consiglio2014}, \cite{leuraud2004}), or counts are available. Fifth, factorial designs can be used, particularly for the interesting interactions are possible as long the number of factors and the number of factor levels is small (due to interpretation, not estimation or computation) \cite{Hothorn2022b}.

%%%%%%%%%%%%%%%%%%%%%%%%%%%%%%%%%%%%%%%%%%%%%%%%%%%%%%%%%%%%%
\subsection{Modifications for variance heterogeneity: the $MCT_{heterogeneity}^E$ trend test}\label{heto}
Specific quantiles for the $\bar{E}_k^2$ test in unbalanced designs with heterogeneous variances were described \cite{miwa2000}, however complex and numerically not available. On the other hand, several publications provide modifications of multiple contrast tests when variance heterogeneity may occur, particularly with the focus of controlling FWER \cite{Hasler2008}, \cite{tamhane2023}, \cite{wen2022}. However, variance heterogeneity, particularly in unbalanced design, may cause also a substantial bias in multiple contrast tests as the Dunnett-type test \cite{Hothorn2023} - a more problematic issue. As a robust modification the use of the sandwich estimator instead of the common means square error estimator \cite{Herberich2012} is considered here. It is particularly easy in R by means of the CRAN packages \textit{multcomp, sandwich}.

%%%%%%%%%%%%%%%%%%%%%%%%%%%%%%%%%%%%%%%%%%%%%%%%%%%%%%%%%%%%%%%%%%%%%%%%%%%%%%%%%%%%%%%%%%%%%%%%
\subsection{Competitors}
Various competitors can be considered, among them the ANOVA F-test for any heterogeneity and the Williams trend test \cite{Williams1971}. The original Williams trend test \cite{Williams1971} is defined for a balanced design (with sample size $n$) as global test: $t_k=(\tilde{y}_k-\bar{y}_0)/\sqrt{(2S^2/n)}$ where $\tilde{y}_k$ is the ML estimator under strict order restriction of the arithmetic means $\bar{y}_i$, e.g. by pooling adjacent violators algorithm (PAVA) \cite{deleuv2009}. Here an approximation by multiple contrast tests \cite{Bretz2006} is used. Both 1-sided and 2-sided tests are considered when available.

%%%%%%%%%%%%%%%%%%%%%%%%%%%%%%%%%%%%%%%%%
\section{A simulation study}
In a simulation study the following 14 tests are compared with respect of empirical size and power under the assumption of Gaussian errors: i) ANOVA F-test, ii) multiple contrast against grand mean; homogeneous variances, 2-sided (MCT2), iii) ibid, heterogeneous variances via sandwich estimator, 2-sided (hetMCT2), iv) homogeneous variances, 2-sided (MCT1), v) heterogeneous variances, 1-sided (hetMCT1), vi) Bartholomew's test, resampling version 
($E^2_k$ ), vii) Williams MCT, homogeneous variances,  2-sided (WIho2), viii) ibid. heterogeneous variances (WIhe2), ix) ibid. 1-sided (WIho1) x) ibid. (WIhe1), xi) new proposed order-restricted multiple contrast test against grand mean, 2-sided, homogeneous variances $(MCT_{ho2}^E$) xii) ibid. heterogeneous variances ($MCT_{he2}^E$), xiii) ibid. 1-sided ($MCT_{ho1}^E$),and xiv) ibid. 1-sided ($MCT_{he1}^E$). The study considered a balanced one-way layout with k=4 with 2500 runs under $H_0$ and 1000 runs under various $H_1$.

\subsection{Monotone shapes in a balanced design with $k=4$ and $n_i=10$ assuming variance homogeneity}
Already in the original paper the power of Williams $t_k$ and $E^2_k$ were compared: $\pi_{t_k}>\pi_{E^2_k}$ for $H_1:\mu_0<\mu_1=...=\mu_k$, but $\pi_{t_k}<\pi_{E^2_k}$ for $H_1:\mu_0=\mu_1=...=\mu_{k-1}<\mu_k$ (particularly with increasing $k$) and $\pi_{t_k}<\pi_{E^2_k}$ for most shapes in-between.

Quantiles of ${E^2_k}$-test are not available for particular unbalanced designs (therefore permutation version in library(Isogene) \cite{otava2017}, this test is less robust for violation of a strict monotonic alternative and simultaneous confidence intervals are not available.\\

\begin{table}[ht]
\centering
\scalebox{0.659}{
\begin{tabular}{r|r|r|rr|rr||r||rrr|r|rr|r>{\columncolor{yellow}}r}
  \hline
Hypo & Profile &  AOV & MCT2 & heMCT2 & MCT1 & heMCT1 & $E^2_k$ & $WI_{ho2}$ & $WI_{he2}$ & $WI_{ho1}$ & $WI_{he1}$& $MCT_{he2}^E$ & $MCT_{ho2}^E$ & $MCT_{he1}^E$ & $MCT_{ho1}^E$ \\ 
  \hline
$H_0$ 	&$\mu_1=\mu_2=\mu_3=\mu_4$  &   0.052 & 0.053 & 0.049 & 0.050 & 0.052 & \Green{0.030} &  0.046 & 0.048 & 0.049 & 0.048 & \Green{0.017} & \Green{0.017} & \Green{0.029} & \Green{0.030} \\ \hline
$H_1$ & $\mu_1=\mu_2=\mu_3<\mu_4$& 0.777 & 0.813 & 0.781 & \textbf{0.853} & 0.823 & 0.803 &  0.701 & 0.695 & 0.823 & 0.789 & 0.768 & 0.819 & 0.831 &  \textbf{0.867}  \\
 & $\mu_1<\mu_2<\mu_3<\mu_4$ &  0.650 & 0.590 & 0.576 & 0.549 & 0.501 & 0.741 &  0.739 & 0.710 & \textbf{0.831} & 0.832 & 0.570 & 0.608 & 0.669 & 0.690  \\  
 & $\mu_1=\mu_2=\mu_3<\mu_4$ &    0.792 & 0.820 & 0.798 & \textbf{0.855} & 0.830 & 0.802 & 0.707 & 0.678 & 0.809 & 0.783 & 0.787 & 0.814 & 0.847 & \textbf{0.869} \\  
& $\mu_1<\mu_2=\mu_3=\mu_4$ &     0.779 & 0.802 & 0.771 & 0.361 & 0.350 & 0.829 &  0.887 & 0.869 & \textbf{0.945} & \textbf{0.941} & 0.771& 0.817 & 0.834 & 0.879  \\ 
& $\mu_1=\mu_2<\mu_3<\mu_4$ &    0.760 & 0.680 & 0.666 & 0.722 & 0.682 & 0.852 & 0.744 & 0.716 & \textbf{0.864} & 0.838 & 0.616 & 0.670 & 0.736 & 0.754 \\ 
& $\mu_1=\mu_2<\mu_3=\mu_4$ &   0.900 & 0.789 & 0.786 & 0.781 & 0.735 & \textbf{0.926} & 0.845 & 0.819 & 0.906 & 0.899 & 0.712 & 0.679 & 0.789 & 0.761 \\ \hline
\end{tabular}
}
\caption{Empirical size and power for variance homogeneity (\Green{conservative}; \textbf{high power})}
\label{tab:ho1}
\end{table}

Table \ref{tab:ho1} illustrates the well-known fact: there is no umpt test for different shapes of monotone alternatives, whereby the two co-competitive tests Williams and new have a relatively high power.
%%%%%%%%%%%%%%%%%%%%%%%%%%%%%%%%%%%%%%%%%%%%%%%%%%%%%%%%%%%%%%%%%
\subsection{Balanced design with $k=4$ and $n_i=10$ for a linear alternative with different non-centralities $\delta$}

\begin{table}[ht]
\centering
\scalebox{0.53}{
\begin{tabular}{rrrrrrrrrrrrrrrrrrrrrrrrrrr}
  \hline
	Hypo & Profile & $\delta$ & AOV & MCT2 & heMCT2 & MCT1 & heMCT1 & E2 & $WI_{ho2}$ & $WI_{he2}$ & $WI_{ho1}$ & $WI_{he1}$& $MCT_{he2}^E$ & $MCT_{ho2}^E$ & $MCT_{he1}^E$ & $MCT_{ho1}^E$&Pit \\ 

$H_1$ & $\mu_1<\mu_2<\mu_3<\mu_4$ &  6 & 0.790 & 0.823 & 0.805 & \textbf{0.857} & 0.833 & 0.802 & 0.709 & 0.685 & 0.819 & 0.783 & 0.778 & 0.814 & 0.841 & \textbf{0.864} &1.08\\  
&  & 5.8 & 0.424 & 0.398 & 0.394 & 0.436 & 0.400 & 0.538 & 0.554 & 0.528 & \textbf{0.674} & 0.649 & 0.399 & 0.422 & 0.512 & 0.542&0.99 \\ 
& & 5.6 & 0.257 & 0.248 & 0.222 & 0.211 & 0.201 & 0.382 & 0.354 & 0.323 & \textbf{0.468} & 0.445 & 0.215 & 0.241 & 0.302 & 0.311& 0.81\\ 
& &5.4 & 0.144 & 0.146 & 0.136 & 0.121 & 0.111 & 0.219 & 0.211 & 0.189 & \textbf{0.289} & 0.268 & 0.120 & 0.129 & 0.170 & 0.190& 0.87\\ 
   \hline
\end{tabular}
}
\caption{Power for decreasing non-centralities (\textbf{high power})}
\label{tab:ho2}
\end{table}

Table \ref{tab:ho2}  shows the complex behavior for different degrees of alternative (i.e. non-centralities): this is also different for various tests. An empirical Pitman efficacy $Pit=\pi_{New}/\pi_{E^2_k}$ shows the quite different relative powers.

%%%%%%%%%%%%%%%%%%%%%%%%%%%%%%%%%%%%%%

\subsection{Monotone shapes in a balanced design with $k=4$ and $n_i=10$ assuming variance heterogeneity}

\begin{table}[ht]
\centering
\scalebox{0.61}{
\begin{tabular}{r|r|r|r|rr|rr||r||rrr|r|rr|rr}
  \hline
Hypo & Profile & $\sigma_i$ & AOV & MCT2 & heMCT2 & MCT1 & heMCT1 & E2 & $WI_{ho2}$ & $WI_{he2}$ & $WI_{ho1}$ & $WI_{he1}$& $MCT_{he2}^E$ & $MCT_{ho2}^E$ & $MCT_{he1}^E$ & $MCT_{ho1}^E$ \\ 
  \hline
$H_0$ &$\mu_1=\mu_2=\mu_3=\mu_4$ & $===\Uparrow$ &  \RED{0.071} & \RED{0.082} & 0.054 & \RED{0.068} & 0.048 & \RED{0.100} 
&\RED{0.073} & 0.049 & \RED{0.062} & 0.048 & \Green{0.018} & \Green{0.041} & \Green{0.027} & 0.057 \\ 
&$\mu_1=\mu_2=\mu_3=\mu_4$ &  $==\Uparrow=$ &  \RED{0.076} & \RED{0.091} & \RED{0.057} & \RED{0.082} & 0.053 & \Green{0.024} &  \Green{0.026} & 0.052 & \Green{0.030} & 0.048 & \Green{0.017} & \Green{0.009} & \Green{0.031} & \Green{0.016}  \\ 
&$\mu_1=\mu_2=\mu_3=\mu_4$ &  $=\Uparrow==$ & \RED{0.080} & \RED{0.096} & \RED{0.070} & \RED{0.066} & \RED{0.062} & 0.034 & \Green{0.014} & \RED{0.060} & \Green{0.022} & 0.042 & \Green{0.012} & \Green{0.006} & \Green{0.022} & \Green{0.014}  \\ 
&$\mu_1=\mu_2=\mu_3=\mu_4$ &  $\Uparrow===$&  \RED{0.100} & \RED{0.114} & \RED{0.060} & \RED{0.094} & \RED{0.074} & \RED{0.090} & \RED{0.176} & \RED{0.076} & \RED{0.110} & \RED{0.064} & \Green{0.016} & 0.056 & 0.042 & \RED{0.072}  \\ 
   \hline \hline

$H_1$ & $\mu_1<\mu_2<\mu_3<\mu_4$& $===\Uparrow$  & \uwave{0.357} & \uwave{0.335} & \uwave{0.345} & \uwave{0.392} & 0.189 & \uwave{0.464} & \uwave{0.459} & 0.450 & \uwave{0.585} & \textbf{0.576} & 0.313 & 0.352 & 0.402 & 0.441  \\ 
 & $\mu_1<\mu_2<\mu_3<\mu_4$ & $==\Uparrow=$ & \uwave{0.365} & \uwave{0.325} & \uwave{0.485} & \uwave{0.349} & 0.395 & 0.514 &  0.451 & 0.600 & 0.600 & \textbf{0.786} & 0.452 & 0.315 & 0.585 & 0.425  \\ 
 & $\mu_1<\mu_2<\mu_3<\mu_4$& $=\Uparrow==$  &  \uwave{0.352} & \uwave{0.342} & \uwave{0.485} & \uwave{0.270} & \uwave{0.391} & 0.462 &  0.428 & \uwave{0.699} & 0.603 & \textbf{0.819} & 0.454 & 0.333 & 0.517 & 0.408  \\ 
 & $\mu_1<\mu_2<\mu_3<\mu_4$& $\Uparrow===$ & \uwave{0.298} & \uwave{0.316} & \uwave{0.304} & \uwave{0.235} & \uwave{0.371} & \uwave{0.444} &  \uwave{0.438} & \uwave{0.274} & \uwave{0.540} & 0.398 & 0.279 & 0.318 & \uwave{0.384} & \uwave{0.379}  \\ 
  \hline \hline
$H_1$ & $\mu_1<\mu_2=\mu_3=\mu_4$& $===\Uparrow$ & \uwave{0.493} & \uwave{0.489} & 0.645 & 0.239 & 0.209 & \uwave{0.510} & \uwave{ 0.580} & 0.722 & \uwave{0.729} & \textbf{0.830} & 0.596 & 0.495 & 0.705 & 0.569  \\ 
& $\mu_1<\mu_2=\mu_3=\mu_4$& $==\Uparrow=$ & \uwave{0.465} & \uwave{0.487} & \uwave{0.650} & \uwave{0.222} & 0.199 & 0.548  & 0.646 & 0.824 & 0.837 & \textbf{0.902} & 0.635 & 0.510 & 0.758 & 0.612 \\
& $\mu_1<\mu_2=\mu_3=\mu_4$& $=\Uparrow==$ & \uwave{0.469} & \uwave{0.478} & \uwave{0.648} & \uwave{0.234} & \uwave{0.214} & 0.562 & 0.796 & 
\uwave{0.814} & 0.801 & \textbf{0.904} & 0.620 & 0.489 & 0.718 & 0.601 \\ 
& $\mu_1<\mu_2=\mu_3=\mu_4$& $\Uparrow===$ & \uwave{0.435} & \uwave{0.479} & \uwave{0.295} & \uwave{0.129} & \uwave{0.234} & \uwave{0.498}  &\uwave{ 0.570} & \uwave{0.356} & \uwave{0.659} & \uwave{0.479} & 0.243 & 0.489 & 0.303 & \uwave{0.555}  \\ 
  \hline
$H_1$ & $\mu_1=\mu_2=\mu_3<\mu_4$& $===\Uparrow$ & \uwave{0.426} & \uwave{0.472} & 0.294 & 0.558 & 0.275 & \uwave{0.518} & \uwave{0.418} & 0.274 & \uwave{0.528} & 0.383 & 0.249 & 0.492 & 0.308 & 0.568 \\ 
 & $\mu_1=\mu_2=\mu_3<\mu_4$& $==\Uparrow=$  & \uwave{0.510} & \uwave{0.540} & \uwave{0.690} & \uwave{0.660} & 0.750 & 0.584 & 0.390 & 0.630 & 0.510 & \textbf{0.760} & 0.660 & 0.520 & 0.780 & 0.660  \\ 
 & $\mu_1=\mu_2=\mu_3<\mu_4$& $=\Uparrow==$  & \uwave{0.480} & \uwave{0.485} & \uwave{0.655} & \uwave{0.572} & \uwave{0.714} & 0.534 &  0.340 & \uwave{0.582} & 0.496 & 0.708 & 0.625 & 0.489 & \textbf{0.729} & 0.601  \\ 
 & $\mu_1=\mu_2=\mu_3<\mu_4$& $\Uparrow===$  & \uwave{0.494} & \uwave{0.505} & \uwave{0.679} & \uwave{0.559} & \uwave{0.705} & \uwave{0.506} & \uwave{0.425} & \uwave{0.269} & \uwave{0.487} & \uwave{0.384} & 0.650 & 0.540 & \textbf{0.720} & \uwave{0.600}  \\ 
  \hline \hline
\end{tabular}
}
\caption{Empirical size and power for variance heterogeneity (\Green{conservative}; \textbf{high power})}
\label{tab:het}
\end{table}

Unsurprisingly, Table \ref{tab:het}  shows clear liberal violations of the test level in the standard tests in the case of different patterns of variance heterogeneity. In other words, these tests are not robust in this respect. The tests with sandwich estimators instead of MQR, on the other hand, are robust \cite{zeileis2020}. Permissible conservative behavior also occurs, mainly caused by the order restriction of the PAVA estimator. Nevertheless, these tests show acceptable power (which is of course reduced compared to the variance-homogeneous case). Again here: no umpt available.
%%%%%%%%%%%%%%%%%%%%%%%%%%%%%%%%%%%%%%%%%%%%%%%%%%%%%%%%%%%%%%%%%%
\subsection{Balanced design with $k=4$ and $n_i=10$ allowing certain non-monotonicity}

\begin{table}[ht]
\centering
\scalebox{0.59}{
\begin{tabular}{r|r|r|r|rr|rr||r||rrr|r|rr|rr}
  \hline
Hypo & Profile & $\sigma_i$ & AOV & MCT2 & heMCT2 & MCT1 & heMCT1 & E2 & $WI_{ho2}$ & $WI_{he2}$ & $WI_{ho1}$ & $WI_{he1}$& $MCT_{he2}^E$ & $MCT_{ho2}^E$ & $MCT_{he1}^E$ & $MCT_{ho1}^E$ \\ 
  \hline
$H_1^{non}$& $\mu_1>\mu_2<\mu_3=\mu_4$ & $====$  & 0.990 & 0.990 & 0.980 & 0.630 & 0.570 & 0.470 & 0.280 & 0.230 & 0.030 & 0.020 & 0.210 & 0.220 & 0.330 & 0.320 \\ \hline \hline 
\end{tabular}
}
\caption{Robustness: power for certain non-monotonic shapes}
\label{tab:non}
\end{table}
Table \ref{tab:non} reveals the remarkable non-robustness of  the order restricted tests for a non-non-monotone shape. 

To summarize the simulations: i) there is no uniformly powerful test (umpt), ii) even the extent of dose-response dependence causes test-specific differences, iii) variance heterogeneity reduces power as expected and changes the size (mostly in a non-permissible liberal direction), but there is an additional bias in the unmodified tests, and iv) the tests are differently non-robust when the monotonicity assumption is violated. As a new test the ($MCT_{he1}^E$) can be proposed for routine evaluation as a robust modification for the Bartholomew trend test.

%%%%%%%%%%%%%%%%%%%%%%%%%%%%%%%%%%%%%%%%%
\section{Analysis of an example data set}
\begin{figure}[htbp]
	\centering
		\includegraphics[width=0.35\textwidth]{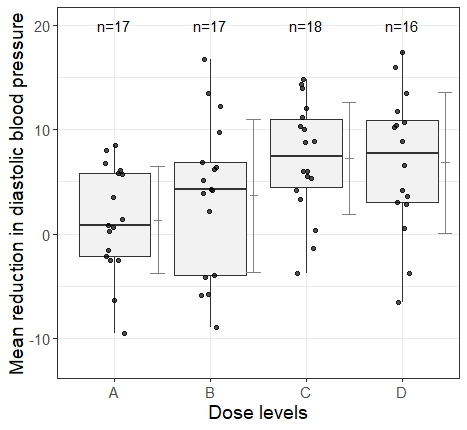}
	\caption{Boxplots of a dose response trial}
	\label{fig:Boxhyp}
\end{figure}

\footnotesize
\begin{verbatim}
library(npordtests) 
library(multcomp); library(sandwich); library(mvtnorm); library(Iso)
data(hypertension)
mod1<-lm(rdbp~doseLevel, data=hypertension) # linear ANOVA-type model
GM<-summary(glht(mod1, linfct = mcp(doseLevel = "GrandMean"), vcov=vcovHC))
ni<-table(hypertension$doseLevel)
MM<-tapply(hypertension$rdbp, hypertension$doseLevel, mean)
SM<-pava(MM,ni); SMM<-mean(MM) # for increasing  weighted pava
BC1<-(SM[1]-SMM);  BC2<-(SM[2]-SMM); BC3<-(SM[3]-SMM); BC4<-(SM[4]-SMM)
bc<-c(BC1, BC2, BC3, BC4) # vector of single contrasts
sig<-GM$test$sigma # MQR estimator
ba<-abs(bc/sig) # pava single studentized contrasts- hetero

k<-nlevels(hypertension$doseLevel)
p.values=c()
for (hh in 1:k){
  p.values[hh]<-1-pmvt(lower =c(rep(-Inf,4)), upper = ba[hh], 
	                corr = cov2cor(GM$vcov), df = GM$df[1])
}
ppp<-min(p.values)
\end{verbatim}
\normalsize

The p-values of the ANOVA F-test for any heterogeneity is $p_{F-test}=0.03$, for the $p_{MCT_{homog 1-sided}^E}=0.011$ and for the 1-sided Williams global test $p_{Wil1}=0.001$.  The smallest p-value for the Williams test is not surprising, as it exclusively uses the large distance between the control (A) and the two upper doses (C+D).

%%%%%%%%%%%%%%%%%%%%%%%%%%%%%%%%%%%%%%%%%%%%%%
\section{Conclusion}
Based on the paradigm 'there is no umpt for ordered alternatives', the test presented is a simple and generalizable version of the original Bartholomew trend test. Corresponding CRAN R software is available. \\
Recently, multiple comparisons procedures for the comparison against a control assuming order restriction was proposed using the using Bartholomew's $\bar{E}_k^2$ test statistics: both for simultaneous confidence intervals and decision trees based on the closed testing procedure \cite{shiraishi2019} (in their chapter 3). The above simulation results reveal that the Williams procedure is a good competitor, easy to calculate and easy for generalization  \cite{Hothorn2020b, Hasler2012, Hothorn2011, Hothorn2010}. 

     \bibliographystyle{apalike}

%			      \bibliography{Barth2024}

\footnotesize
			\begin{thebibliography}{}

\bibitem[Armitage, 1955]{Armitage1955}
Armitage, P. (1955).
\newblock Tests for linear trends in proportions and frequencies.
\newblock {\em Biometrics}, 11(3):375--386.

\bibitem[Bartholomew, 1959]{bartholomew1959}
Bartholomew, D. (1959).
\newblock A test of homogeneity for ordered alternatives.
\newblock {\em Biometrika}, 46(1/2):36--48.

\bibitem[Bretz, 2006]{Bretz2006}
Bretz, F. (2006).
\newblock An extension of the Williams trend test to general unbalanced linear
  models.
\newblock {\em Computational Statistics and Data Analysis, vol. 50, no. 7, Art.
  no. 7, 2006.}

\bibitem[Consiglio et~al., 2014]{consiglio2014}
Consiglio, J., Shan, G., and Wilding, G. (2014).
\newblock A comparison of exact tests for trend with binary endpoints using
  Bartholomew statistic.
\newblock {\em The International Journal of Biostatistics}, 10(2):221--230.

\bibitem[De~Leeuw et~al., 2009]{deleuv2009}
De~Leeuw, J., Hornik, K., and Mair, P. (2009).
\newblock Isotone optimization in R: pool-adjacent-violators algorithm (pava)
  and active set methods.
\newblock {\em Journal of Statistical Software}, 32(5):1--24.

\bibitem[Hasler and Hothorn, 2008]{Hasler2008}
Hasler, M. and Hothorn, L.~A. (2008).
\newblock Multiple contrast tests in the presence of heteroscedasticity.
\newblock {\em Biometrical Journal}, 50(5):793--800.

\bibitem[Hasler and Hothorn, 2012]{Hasler2012}
Hasler, M. and Hothorn, L.~A. (2012).
\newblock A multivariate williams-type trend procedure.
\newblock {\em Statistics in Biopharmaceutical Research}, 4(1):57--65.

\bibitem[Herberich and Hothorn, 2012]{Herberich2012}
Herberich, E. and Hothorn, L.~A. (2012).
\newblock Statistical evaluation of mortality in long-term carcinogenicity
  bioassays using a Williams-type procedure.
\newblock {\em Regulatory Toxicology and Pharmacology}, 64(1):26--34.

\bibitem[Hothorn, 2020]{Hothorn2020b}
Hothorn, L. (2020).
\newblock Comparisons of proportions in k dose groups against anegative control
  assuming order restriction: Williams-type test vs. closed test procedures.
\newblock {\em arXiv 2011.13758v1}.

\bibitem[Hothorn, 2022]{Hothorn2022b}
Hothorn, L.~A. (2022).
\newblock Simultaneous confidence intervals for the interpretation of primary
  and secondary effects in factorial designs without a pre-test on interaction.
\newblock {\em arXiv preprint arXiv:2204.08336}.

\bibitem[Hothorn and Djira, 2011]{Hothorn2011}
Hothorn, L.~A. and Djira, G.~D. (2011).
\newblock A ratio-to-control williams-type test for trend.
\newblock {\em Pharmaceutical Statistics}, 10(4):289--292.

\bibitem[Hothorn and Hasler, 2023]{Hothorn2023}
Hothorn, L.~A. and Hasler, M. (2023).
\newblock The Dunnett procedure with possibly heterogeneous variances.
\newblock {\em arXiv preprint arXiv:2303.09222}.

\bibitem[Hothorn et~al., 2010]{Hothorn2010}
Hothorn, L.~A., Sill, M., and Schaarschmidt, F. (2010).
\newblock Evaluation of incidence rates in pre-clinical studies using a
  Williams-type procedure.
\newblock {\em International Journal of Biostatistics}, 6(1):15.

\bibitem[Hothorn et~al., 2008]{Hothorn2008}
Hothorn, T., Bretz, F., and Westfall, P. (2008).
\newblock Simultaneous inference in general parametric models.
\newblock {\em Biometrical Journal}, 50(3):346--363.

\bibitem[Konietschke et~al., 2013]{Konietschke2013}
Konietschke, F., Bosiger, S., Brunner, E., and Hothorn, L.~A. (2013).
\newblock Are multiple contrast tests superior to the anova?
\newblock {\em International Journal of Biostatistics}, 9(1):63--73.

\bibitem[Lei et~al., 1995]{lei1995}
Lei, X., Peng, Y., and Wright, F. (1995).
\newblock Testing homogeneity of normal means with a simply ordered alternative
  and dependent observations.
\newblock {\em Computational Statistics \& Data Analysis}, 20(2):173--183.

\bibitem[Leuraud and Benichou, 2004]{leuraud2004}
Leuraud, K. and Benichou, J. (2004).
\newblock Tests for monotonic trend from case-control data:
  Cochran-Armitage-Mantel trend test, isotonic regression and single and
  multiple contrast tests.
\newblock {\em Biometrical Journal}, 46(6):731--749.

\bibitem[Lin et~al., 2015]{Lin2015}
Lin, D., Pramana, S., Verbeke, T., and Shkedy, Z. (2015).
\newblock {\em IsoGene: Order-Restricted Inference for Microarray Experiments}.
\newblock R package version 1.0-24.

\bibitem[Miwa, 1998]{miwa1998}
Miwa, T. (1998).
\newblock Bartholomw's test as a multiple contrast test and its applications.
\newblock {\em Jap. J. Biometr.}, 19(1):1--9.

\bibitem[Miwa et~al., 2000]{miwa2000}
Miwa, T., Hayter, A., and Liu, W. (2000).
\newblock Calculations of level probabilities for normal random variables with
  unequal variances with applications to Bartholomew s test in unbalanced
  one-way models.
\newblock {\em Computational Statistics \& Data Analysis}, 34(1):17--32.

\bibitem[Otava et~al., 2017]{otava2017}
Otava, M., Sengupta, R., Shkedy, Z., Lin, D., Pramana, S., Verbeke, T.,
  Haldermans, P., Hothorn, L.~A., Gerhard, D., Kuiper, R.~M., Klinglmueller,
  F., and Kasim, A. (2017).
\newblock Isogenegui: Multiple approaches for dose-response analysis of
  microarray data using R.
\newblock {\em R Journal}, 9(1):14--26.

\bibitem[Pallmann and Hothorn, 2016]{Pallmann2016}
Pallmann, P. and Hothorn, L.~A. (2016).
\newblock Analysis of means: a generalized approach using R.
\newblock {\em Journal of Applied Statistics}, 43(8):1541--1560.

\bibitem[Shiraishi et~al., 2019]{shiraishi2019}
Shiraishi, T.-a., Sugiura, H., Matsuda, S.-i., et~al. (2019).
\newblock {\em Pairwise multiple comparisons: Theory and computation}.
\newblock Springer.

\bibitem[Tamhane and Xi, 2023]{tamhane2023}
Tamhane, A.~C. and Xi, D. (2023).
\newblock Multiplicity adjustments for the dunnett procedure under
  heterokcedasticity.
\newblock {\em Biometrical Journal}, page 2200300.

\bibitem[Turner, 2023]{Turner2023}
Turner, R. (2023).
\newblock {\em Iso: Functions to Perform Isotonic Regression}.
\newblock R package version 0.0-21.

\bibitem[Wen et~al., 2022]{wen2022}
Wen, M.-J., Wen, C.-C., and Wang, W.-M. (2022).
\newblock Single-stage sampling procedure for heteroscedasticity in multiple
  comparisons with a control.
\newblock {\em Communications in Statistics-Simulation and Computation}, pages
  1--10.

\bibitem[Williams, 1971]{Williams1971}
Williams, D.~A. (1971).
\newblock Test for differences between treatment means when several dose levels
  are compared with a zero dose control.
\newblock {\em Biometrics}, 27(1):103f.

\bibitem[Zeileis et~al., 2020]{zeileis2020}
Zeileis, A., K{\"o}ll, S., and Graham, N. (2020).
\newblock Various versatile variances: an object-oriented implementation of
  clustered covariances in R.
\newblock {\em Journal of Statistical Software}, 95:1--36.

\end{thebibliography}

\end{document}